\documentclass[12pt,a4paper]{article}
\usepackage[latin9]{inputenc}
\usepackage{amsmath}
\usepackage{amssymb}
\usepackage{graphicx}
\usepackage{esint}

\makeatletter

\pdfpageheight\paperheight
\pdfpagewidth\paperwidth

\usepackage{jheppub}%\usepackage{axodraw4j}

\author[a,b]{R.N. Lee}
\affiliation[a]{Budker Institute of Nuclear Physics, 630090, Novosibirsk, Russia}
\affiliation[b]{Novosibirsk State University, 630090, Novosibirsk, Russia}
\emailAdd{R.N.Lee@inp.nsk.su}
\dedicated{Dedicated to my father`s birthday}
\makeatother

\begin{document}
\global\long\def\Sum{\mathop{\Sigma}}
 \abstract

\global\long\def\d{d}

\abstract{\emph{Mathematica }package \texttt{LiteRed} is described.
It performs the heuristic search of the symbolic IBP reduction rules
for loop integrals. It implements also several convenient tools for
the search of the symmetry relations, construction of the differential
equations and dimensional recurrence relations.}

\title{Presenting \texttt{LiteRed}: a tool for the Loop InTEgrals REDuction}
\maketitle
\allowdisplaybreaks

\section{Introduction}

Demand for the multiloop calculations is constantly growing nowadays.
As the number of loops increases, the calculations become more and
more complicated and require, almost necessarily, some stages to be
done automatically using computers. The reduction of the loop integrals
with the help of the integration-by-parts (IBP) identities \cite{ChetTka1981,Tkachov1981}
is an important stage which can be automatized. This reduction allows
one to reduce the calculation of the multiloop diagrams to the calculation
of a finite set of master integrals. It is important that the reduction
also allows one to obtain differential and difference equations, which
can be used for the calculation of the loop integrals without explicit
itegration.

One of the most successful methods of the IBP reduction is the Laporta
algorithm \cite{Laporta2000}. The algorithm is easy to implement
and to use, given a sufficient amount of time, it works flawlessly.
It also allows for a number of programming improvements. These advantages
explain why many modern most powerful reduction programs heavily rely
on this algorithm, in particular,\texttt{ AIR} \cite{AnastasiouLazopoulos2004},
\texttt{FIRE} \cite{Smirnov2008}, \texttt{Reduze} \cite{Studeru2010,ManteuffelStuderus2012},
and many private versions. However, there are some weak points of
this algorithm, which may put some restrictions on its application.
First, the reduction generates heavy-weight databases of the discovered
rules which can be too expensive to save. Therefore, typically, the
reduction is performed each time ``from the beginning'' which requires
the same identities to be solved in each run. Another disadvantage
of the Laporta algorithm is due to the huge redundancy of the IBP
identities \cite{Lee2008}. This redundancy results in many unnecessary
calculations which make the reduction slow.

Another approach to the reduction is a derivation of the symbolic
rules which can be applied to any problem of a given class. Its advantages
are obvious: nothing is being solved in the process of reduction,
therefore, the reduction is very fast. Symbolic rules are small in
size, so, they can be easily saved for future calculations. The bottleneck
of the approach is the search of the symbolic rules, a stage which
seems to require a lot of manual work.

An ideal solution of the reduction problem would be, therefore, designing
a program which automatically finds the symbolic reduction rules at
the first stage and then uses those rules for the reduction. Much
effort has been devoted to the developement of the approach connected
with the notion of the Groebner basis \cite{Tarasov1998,Gerdt2005a,SmirSmi2006a,Smirnov:2006tz}.
This is due to the fact that the problem of IBP reduction is very
similar to the reduction of the elements of some algebra with respect
to its ideal. In the latter problem there is a known algorithm of
the construction of the Groebner basis --- the Buchberger's algorithm
and its generalizations. When the Groebner basis is constructed, the
reduction can be performed unambiguously and very fast. If this approach
always worked for the IBP reduction, it would definitely be the most
complete solution of the reduction problem. However, there is a small
peculiarity of the IBP identities which results in the essential obstacle
when generalizing the Buchberger's algorithm to the IBP reduction.
As it was shown in Ref. \cite{Lee2008}, the IBP reduction problem
can be reduced to the problem of reduction of some noncommuting polynomial
ring with respect to the direct sum of the left and right ideals.
While the Buchberger's algorithm seperately works for both left and
right ideals, it appears to be difficult to generalize this algorithm
to the desired case. Probably, the only, partly successful, attempt
of this approach has been made in \texttt{FIRE}, where the notion
of s-bases \cite{SmirSmi2006a,Smirnov:2006tz} have been used.

Of course, there are many peculiar features of the IBP identities
which make the IBP reduction not the general-case reduction with respect
to the direct sum of the left and right ideals. In particular, as
it was shown in Ref. \cite{Lee2008}, the generating set of the left
ideal appears to be equipped with a Lie-algebraic structure, and that
of the right ideal consists of the commuting elements. It is quite
possible that at some point in the future a general solution of the
IBP reduction problem will appear. Meanwhile, one can try to develop
some heuristic agorithms which are not guaranteed to work for each
case, but, nevertheless, are useful from the practical point of view.
With the lack of a systematic approach (i.e., a strict algorithm),
this developement can be quite challenging. 

This short note describes a \textsl{Mathematica} package \texttt{LiteRed}
which can be considered as an attempt of the implementation of the
heuristic approach to the IBP reduction. The package can be downloaded
as a zip archive from 

\texttt{http://www.inp.nsk.su/\textasciitilde{}lee/programs/LiteRed/}

\section{General setup}

Assume that we are interested in the calculation of the $L$-loop
integral depending on the $E$ external momenta $p_{1},\ldots,p_{E}$.
There are $N=L(L+1)/2+LE$ scalar products depending on the loop momenta
$l_{i}$:
\begin{equation}
s_{ij}=l_{i}\cdot q_{j}\,,\ 1\leqslant i\leqslant L,j\leqslant L+E,
\end{equation}
where $q_{1,\ldots,L}=l_{1,\ldots,L}$, $q_{L+1,\ldots,L+E}=p_{1,\ldots,E}$.

The loop integral has the form
\begin{equation}
J\left(\boldsymbol{n}\right)=J(n_{1},n_{2},\ldots,n_{N})=\int d^{\d}l_{1}\ldots d^{\d}l_{L}j(\boldsymbol{n})=\int\frac{d^{\d}l_{1}\ldots d^{\d}l_{L}}{D_{1}^{n_{1}}D_{2}^{n_{2}}\ldots D_{N}^{n_{N}}}\,,\label{eq:J}
\end{equation}
where the scalar functions $D_{\alpha}$ are linear polynomials with
respect to $s_{ij}$. The functions $D_{\alpha}$ are assumed to be
linearly independent and to form a complete basis in the sense that
any non-zero linear combination of them depends on the loop momenta,
and any $s_{ik}$ can be expressed in terms of $D_{\alpha}$. Thus,
each integral is associated with a point in $\mathbb{Z}^{N}$. Some
of the functions $D_{\alpha}$ correspond to the denominators of the
propagators, the other correspond to the irreducible numerators. E.g.,
the $K$-legged $L$-loop diagram corresponds to $E=K-1$ and the
maximal number of denominators is $M=E+3L-2$, so that the rest $N-M=(L-1)(L+2E-4)/2$
functions correspond to irreducible numerators.

\paragraph{IBP identities}

The IBP identities \cite{ChetTka1981,Tkachov1981} are based on the
fact that, in the dimensional regularization, the integral of the
total derivative is zero. They are derived from the identity
\begin{equation}
0=\int d^{\d}l_{1}\ldots d^{\d}l_{L}\frac{\partial}{\partial l_{i}}\cdot q_{k}j(\boldsymbol{n})\,.\label{eq:IBP}
\end{equation}
Performing the differentiation on the right-hand side and expressing
the scalar products via $D_{\alpha}$, we obtain the recurrence relation
for the function $J$.

\paragraph{LI identities}

There is also another class of identities, called Lorentz-invariance
(LI) identities due to the fact that the integral (\ref{eq:J}) is
Lorentz scalar \cite{GehrRem2000}. They have the form
\begin{equation}
p_{i}^{\mu}p_{j}^{\nu}\left(\sum_{k}p_{k[\nu}\frac{\partial}{\partial p_{k}^{\mu]}}\right)J(n_{1},n_{2},\ldots,n_{N})=0\,.\label{eq:LI}
\end{equation}

The differential operator in braces is nothing but the generator of
the Lorentz transformation in the linear space of scalar functions
depending on $p_{k}$. Again, performing the differentiation on the
right-hand side, we obtain LI identity. Though these identities can
be represented as a linear combination of the IBP identities \cite{Lee2008},
they prove to be useful in real-life reduction.

\paragraph{Sectors}

The notion of sectors can be introduced as follows. The $\boldsymbol{\theta}=\left(\theta_{1},\ldots,\theta_{N}\right)$
\emph{sector}, where $\theta_{i}=0,1$, is a set of all points $\left(n_{1},\ldots,n_{N}\right)$
in $\mathbb{Z}^{N}$ whose coordinates obey the condition 
\begin{align}
\Theta\left(n_{\alpha}-1/2\right) & =\theta_{\alpha}
\end{align}
In particular, the point $(\theta_{1},\ldots,\theta_{N})$ belongs
to the $(\theta_{1},\ldots,\theta_{N})$ sector, and will be referred
to as the \emph{corner point of the sector}. Owing to this definition,
the integrals of the same sector have the same set of denominators.

\paragraph{Scaleless integrals}

The scaleless integral can be defined as the one which gains additional
non-unity factor under some linear transformation of the loop momenta.
Obviously, if $j\left(\theta_{1},\ldots,\theta_{N}\right)$ is scaleless,
then all integrals of the sector $\left(\theta_{1},\ldots,\theta_{N}\right)$
are zero. We will call such a sector a \emph{zero sector}. A simple
and convenient criterion of zero sectors has been formulated in Ref.
\cite{Lee2008}. According to this criterion, the sector is zero if
the solution of the IBP equations in the corner point $\left(\theta_{1},\ldots,\theta_{N}\right)$
result in the identity $ $$j\left(\theta_{1},\ldots,\theta_{N}\right)=0$.
Note that this criterion may miss some scaleless sectors. Though this
seems to be a  small problem (undetected zero sectors are simply reduced
to lower sectors), let us explain on a simple example why this happens.

Let us consider the massless one-loop onshell propagator integral
\[
J\left(n_{1},n_{2}\right)=\int\frac{d^{\d}l}{\left[l^{2}\right]^{n_{1}}\left[\left(l-k\right)^{2}\right]^{n_{2}}},\quad k^{2}=0\,.
\]

Obviously, this integral is zero for any $n_{1}$ and $n_{2}$. However,
it can be explicitely checked that there is no linear combination
of the operators $\partial_{l}\cdot l$ and $\partial_{l}\cdot k$
which acts as identity on the integrand for $n_{1}=n_{2}=1$. In other
words, the solution of the IBP identities in the corner point of the
sector $\left(1,1\right)$ does not result directly to $J\left(1,1\right)=0$
(though, it results to, e.g., $J\left(1,1\right)\propto J\left(0,2\right)$
). In order to prove that the integral $J\left(1,1\right)$ is scaleless,
let us consider instead the following operator
\[
O=\partial_{l}\cdot\left(l+\left(l\cdot k\right)\tilde{k}-\left(l\cdot\tilde{k}\right)k\right),
\]
where $\tilde{k}$ is an auxiliary vector chosen to satisfy the conditions
$\tilde{k}^{2}=0$ and $\tilde{k}\cdot k=1$. It is easy to check
that $Oj\left(1,1\right)=\left(\d-4\right)j\left(1,1\right)$. Since
the operator $O$ is a generator of the linear transformation $l\to l+\epsilon\left(l+\left(l\cdot k\right)\tilde{k}-\left(l\cdot\tilde{k}\right)k\right)$,
the integral $j\left(1,1\right)$ is scaleless. The reason why the
IBP identities failed to lead to the identity $J\left(1,1\right)=0$
is that the construction of this identity required introduction of
the auxiliary vector $\tilde{k}$. There is an interesting open question:
whether the introduction of the auxiliary vectors (or tensors) may
lead to a new kind of the identities independent of IBP identities
(unlikely) or dependent on them, but still useful for real-life reduction.

\paragraph{Symmetry relations}

In many cases there exist nontrivial linear transformations of the
loop momenta which map the set of the denominators of one sector on
the set of the denominators of the same, or another, sector. Those
transformations have the form 
\[
l_{i}\to M_{ij}q_{j},
\]
where $M_{ij}$ is a $L\times\left(L+E\right)$ matrix. It is easy
to understand that, for nonzero sectors, there is only a finite number
of such transformations, all subjected to the condition $\left|\det\left\{ \left.M_{ij}\right|_{i,j=1,\ldots L}\right\} \right|=1$.
Those transformations induce some mappings of the denominator set
and also mappings of the numerators into the linear polynomials of
$D_{i}$ . These mappings give nontrivial identities between the integrals
of two different (or one) sectors, which are conventionally called
\emph{symmetry relations} (SR).

\paragraph{Ordering}

For the reduction procedure to work, it is necessary to define some
suitable ordering of the integrals, i.e., the ordering in $\mathbb{Z}^{N}$.
It is natural to consider the integrals with less denominators to
be simpler. This defines a partial ordering of the sectors. We can
extend this ordering to the complete one by, e.g., saying, that the
two sectors $\boldsymbol{\theta}_{1}$ and $\boldsymbol{\theta}_{2}$
with equal number of denominators are ordered lexicographically. For
the integrals in sector $ $$\boldsymbol{\theta}=(\theta_{1},\ldots,\theta_{N})$
with $K=\sum_{\alpha=1}^{N}\Theta\left(n_{\alpha}-1/2\right)$ denominators
we choose the folowing ordering. For the integral $J\left(n_{1},\ldots,n_{N}\right)$
we determine the ordering weight $\mathbf{w}=\left(w_{-1},\, w_{0},\, w_{1},\ldots,\, w_{N}\right)$,
where $w_{-1}=\sum_{\alpha=1}^{N}|n_{\alpha}|$ is a total power of
the denominators and numerators, $w_{0}=\sum_{\alpha=1}^{N}\left(1-\theta_{\alpha}\right)n_{\alpha}$
is the total power of the numerators, and $\left(w_{1},\ldots,w_{N}\right)$
is obtained from $\left(-\left|n_{1}\right|,\ldots,-\left|n_{N}\right|\right)$
by shifting powers, corresponding to denominators, to $K$ left-most
positions. Then, the two integrals in one sector are compared by comparing
the left-most dictinct entries of their ordering weights. In particular,
the simplest integral in the sector $(\theta_{1},\ldots,\theta_{N})$
is $J(\theta_{1},\ldots,\theta_{N})$.

Of course, the choice of the ordering is not unique. In principle,
for the possibility of the reduction, there is only one strictly required
property of the ordering. It is that for any integral $J\left(\mathbf{n}\right)$
there is only finite number of the integrals simpler than it (this
number depends, of course, on $\mathbf{n}$). However, the following
condition of the chosen ordering is essential for our consideration.
For a given sector all components of the ordering weight $\mathbf{w}$
are just linear combinations of $n_{i}$. In particular, it leads
to the fact that the relation $J\left(\mathbf{n}_{1}\right)\prec J\left(\mathbf{n}_{2}\right)$
between the integrals in the same sector is invariant with respect
to the shift of both $\mathbf{n}_{1}$ and $\mathbf{n}_{2}$ by some
$\delta\mathbf{n}$, provided that $J\left(\mathbf{n}_{1}+\delta\mathbf{n}\right)$
and $J\left(\mathbf{n}_{2}+\delta\mathbf{n}\right)$ also belong to
the same sector.

\paragraph{Differential equations}

As it was mentioned above, the differential equations can be used
for finding the master integrals. The simplest type of such equations
is the differential equation with respect to the mass. Probably, the
first example of their application is presented in Refs. \cite{Kotikov1991,Kotikov1991a,Kotikov1991b}.
The differential equations with respect to the invariant constructed
of the external momenta have been introduced and applied in Refs.
\cite{GehrRem2000,GehrRem2001,GehrRem2001a}. The peculiarity of the
latter case is due to the fact that, though the \emph{integral} depends
on the external momenta only via their scalar products, the \emph{integrand}
also depends on the scalar products of the external momenta and loop
momenta. Therefore, before differentiating the integral, it is necessary
to express the derivative with respect to the invariant via the derivatives
with respect to the external momenta. For example, if the integral
depends on $p^{2}$, $q^{2}$ and $p\cdot q$, the derivative with
respect to $p\cdot q$ at fixed $p^{2}$ and $q^{2}$ can be expressed
in two equivalent ways
\[
\frac{\partial}{\partial\left(p\cdot q\right)}J\left(\mathbf{n}\right)=\frac{\left(p\cdot q\right)p-p^{2}q}{\left(p\cdot q\right)^{2}-p^{2}q^{2}}\cdot\frac{\partial}{\partial p}J\left(\mathbf{n}\right)=\frac{\left(p\cdot q\right)q-q^{2}p}{\left(p\cdot q\right)^{2}-p^{2}q^{2}}\cdot\frac{\partial}{\partial q}J\left(\mathbf{n}\right)\,.
\]
In general case, when there are $E>2$ external vectors, we have the
following formulas:
\begin{align}
\frac{\partial}{\partial\left(p_{1}\cdot p_{2}\right)}J\left(\mathbf{n}\right) & =\sum\left[\mathbb{G}^{-1}\right]_{i2}p_{i}\cdot\partial_{p_{1}}J\left(\mathbf{n}\right)=\sum\left[\mathbb{G}^{-1}\right]_{i1}p_{i}\cdot\partial_{p_{2}}J\left(\mathbf{n}\right)\,,\nonumber \\
\frac{\partial}{\partial\left(p_{1}^{2}\right)}J\left(\mathbf{n}\right) & =\frac{1}{2}\sum\left[\mathbb{G}^{-1}\right]_{i1}p_{i}\cdot\partial_{p_{1}}J\left(\mathbf{n}\right)\,.\label{eq:Dinv}
\end{align}
where $\mathbb{G}=\mathbb{G}\left(p_{1},\dots,p_{E}\right)=\begin{pmatrix}p_{1}^{2} & \cdots & p_{1}\cdot p_{E}\\
\vdots & \ddots & \vdots\\
p_{1}\cdot p_{E} & \cdots & p_{E}^{2}
\end{pmatrix}$ is a Gram matrix.

Acting by the operator on the right-hand side on the integrand and
performing the IBP reduction, one obtains the differential equation
for $J\left(\mathbf{n}\right)$.

\paragraph{Dimensional recurrences }

Dimensional recurrences have been introduced in Ref. \cite{Tarasov1996}.
Since their introduction, they have been successfully applied to the
calcultion of the different integrals, see Refs. \cite{Tarasov1996,Tarasov2000,Tarasov2006}.
Recently, a method of calculation of the multiloop master integrals,
based on the dimensional recurrences and analytical properties of
the multiloop integrals as functions of $\d$, has been introduced
in Ref. \cite{Lee2010} and successfully applied in Refs. \cite{LeeSmSm2010,LeeSmi2010,LeeSmSm2010a,Lee2010a,LeeTer2010,LeeSmSm2011,Lee2011e,LeeSmirnov2012}.
For further purposes it is convenient to introduce the operators $A_{\alpha}$
and $B_{\alpha}$, see Ref. \cite{Lee2008}, acting as follows
\begin{align}
\left(A_{i}J^{\left(\mathcal{D}\right)}\right)\left(n_{1},\ldots,n_{N}\right) & =n_{i}J^{\left(\mathcal{D}\right)}\left(n_{1},\ldots,n_{i}+1,\ldots,n_{N}\right),\nonumber \\
\left(B_{i}J^{\left(\mathcal{D}\right)}\right)\left(n_{1},\ldots,n_{N}\right) & =J^{\left(\mathcal{D}\right)}\left(n_{1},\ldots,n_{i}-1,\ldots,n_{N}\right).
\end{align}

The original derivation of the dimensional recurrence relation in
Ref. \cite{Tarasov1996} relied on the parametric representation.
The result of this derivation for the integrals which can be presented
as a graph has a nice and compact form
\begin{equation}
J^{\left(\d-2\right)}\left(\mathbf{n}\right)=\mu^{L}\sum_{\text{trees}}\left(A_{i_{1}}\ldots A_{i_{L}}J^{\left(\d\right)}\right)\left(\mathbf{n}\right),\label{eq:rDRRo}
\end{equation}
where $i_{1},\ldots,i_{L}$ numerate the chords of the tree, and $\mu=\pm1$
for the Euclidean/Minkovskian case, respectively. For computer implementations,
probably, a more convenient formula has been derived in Ref. \cite{Lee2010a}.
It reads
\begin{equation}
J^{\left(\d-2\right)}\left(\mathbf{n}\right)=(\mu/2)^{L}\det\left\{ 2^{\delta_{ij}}\frac{\partial D_{k}}{\partial s_{ij}}A_{k}|_{i,j=1\ldots L}\right\} J^{\left(\d\right)}\left(\mathbf{n}\right).\label{eq:rDRR}
\end{equation}
Note that this formula is valid also for the integrals with numerators
and dots. We will call Eqs. \eqref{eq:rDRRo} and \eqref{eq:rDRR}
the \emph{raising dimensional recurrence relation}s.

For completeness, we present also the \emph{lowering dimensional recurrence
relation}, obtained in Ref. \cite{Lee2010}

\begin{equation}
J^{\left(\d+2\right)}\left(\mathbf{n}\right)=\frac{(2\mu)^{L}\left[V\left(p_{1},\ldots,p_{E}\right)\right]^{-1}}{\left(\d-E-L+1\right)_{L}}P\left(B_{1},\ldots,B_{N}\right)J^{\left(\d\right)}\left(\mathbf{n}\right),\label{eq:lDRR}
\end{equation}
where $\alpha_{L}=\alpha\left(\alpha+1\right)\ldots\left(\alpha+L-1\right)$
is the Pochhammer symbol, $V\left(v_{1},\ldots v_{k}\right)=\det\mathbb{G}\left(v_{1},\ldots v_{k}\right)$
is the Gram determinant, and $P\left(D_{1},\ldots,D_{N}\right)=V(q_{1},\ldots q_{L+E})$.

If we consider one of these dimensional recurrence relations for the
master integral and make the IBP reduction of the right-hand side,
we will obtain the difference equation for this master integral.

\section{Using the \texttt{LiteRed} package}

\subparagraph{}

\begin{figure}
\includegraphics[height=1\textheight]{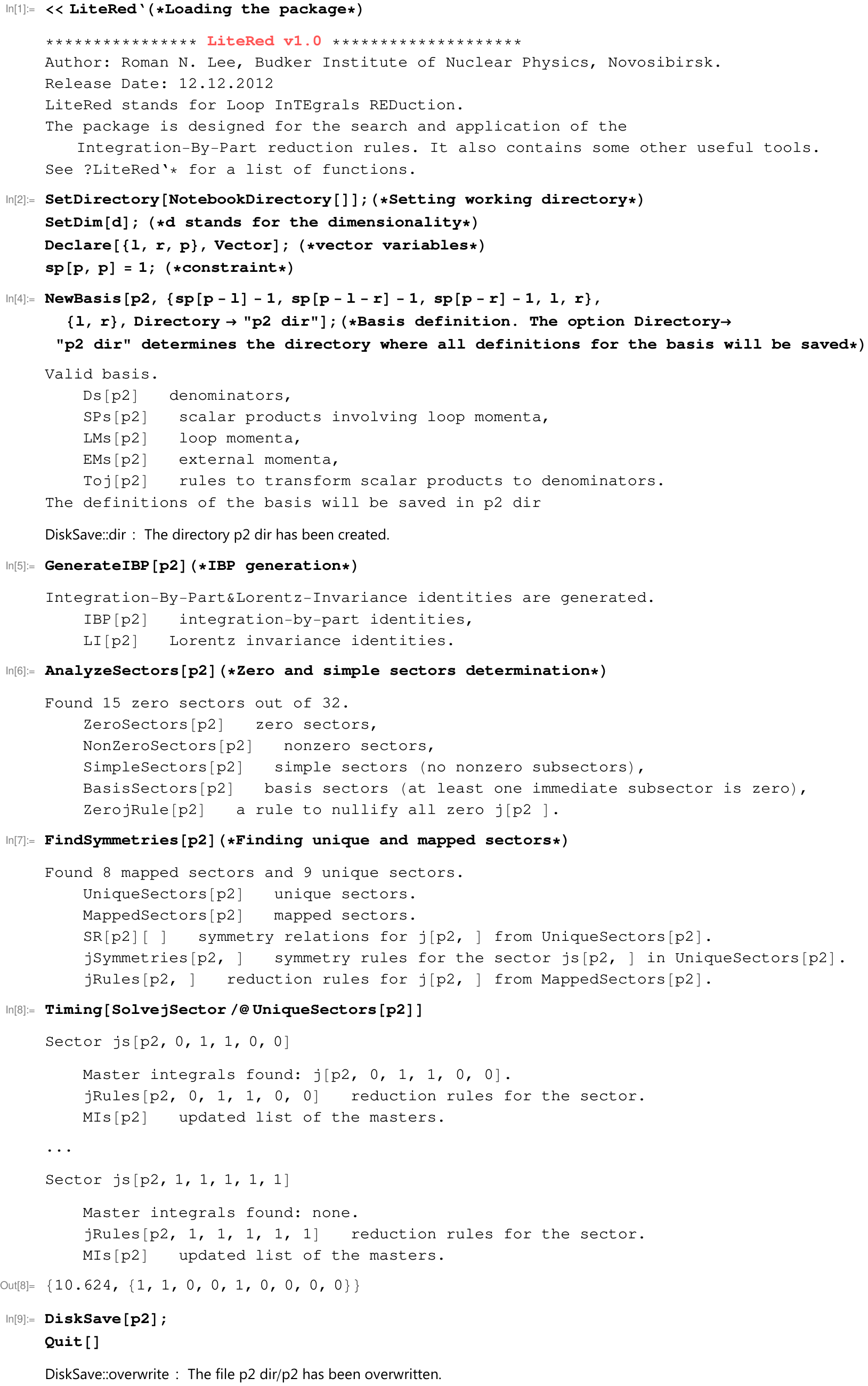}\caption{A simple program of finding IBP rules.\label{fig:A-simple-program}}
\end{figure}

A typical package usage includes two stages: the search of the reduction
rules and their application for the reduction. During the first stage
the definitions, related to the basis, in particular, the reduction
rules, can be saved to the disk. These definitions should be loaded
 on the second stage and applied for the reduction. The basic example
of a program searching the reduction rules is presented in Fig. \ref{fig:A-simple-program}.
This program finds the reduction rules for the two-loop onshell massive
propagator. 

Let us provide some comments. At each stage the program generates
some objects which are then used at the later stages. E.g., the command
\linebreak \texttt{\textbf{NewBasis{[}p2,\{sp{[}p-l{]}-1,sp{[}p-l-r{]}-1,sp{[}p-r{]}-1,l,r\},\{l,r\}{]}}}
sets up a new basis \texttt{\textbf{p2}}, consisting of the functions
$D_{\alpha}$ depending linearly on the scalar products involving
loop momenta \texttt{\textbf{l}}\textbf{, }\texttt{\textbf{r}} (\texttt{\textbf{sp{[}a,b{]}}}
stands for the scalar product of \texttt{\textbf{a}} and \texttt{\textbf{b}},
\texttt{\textbf{sp{[}a{]}}} is a shortcut for \texttt{\textbf{sp{[}a,a{]}}}).
This procedure checks that the set of $D_{\alpha}$ is linearly independent
and complete and generates several objects \texttt{\textbf{Ds{[}p2{]}}}\texttt{,
}\texttt{\textbf{SPs{[}p2{]}}}\texttt{, }\texttt{\textbf{LMs{[}p2{]}}}\texttt{,
}\texttt{\textbf{EMs{[}p2{]}}}\texttt{,}\texttt{\textbf{ Toj{[}p2{]}}}.
The meaning of these objects should be clear from the output immediately
following the\textbf{ }\texttt{\textbf{NewBasis}} command. The objects
\texttt{\textbf{IBP{[}p2{]}}} and \texttt{\textbf{LI{[}p2{]}}} generated
by \texttt{\textbf{GenerateIBP}}\textbf{ }call give the functions
which return the Integration-By-Parts identities and Lorentz-Invariance
identities in a given point, see Eqs. \eqref{eq:IBP} and \eqref{eq:LI}.
E.g., \texttt{\textbf{IBP{[}p2{]}{[}n1,n2,n3,n4,n5{]}}} gives a list
of IBP identities in a general point. The IBP identities generated
by \texttt{\textbf{GenerateIBP}} procedure are necessary for the detemination
of zero sectors by the \texttt{\textbf{AnalyzeSectors}} procedure.
This procedure generates the list \texttt{\textbf{ZeroSectors{[}p2{]}}}
of zero sectors as well as the list \texttt{\textbf{SimpleSectors{[}p2{]}}}
of the simplest nonzero sectors. The latter is used in the procedure
\texttt{\textbf{FindSymmetries}}. This procedure finds internal and
mutual symmetries of the nonzero sectors and generates the lists \texttt{\textbf{MappedSectors{[}p2{]}}}
and \texttt{\textbf{UniqueSectors{[}p2{]}}}. Any sector from the former
list can be mapped onto some sector from the latter. The substitution
rules for such a mapping can be found in \texttt{\textbf{jRules{[}p2,...{]}}}
for each sector \texttt{\textbf{js{[}p2,...{]}}} from the list \texttt{\textbf{MappedSectors{[}p2{]}}}. 

\begin{figure}
\includegraphics{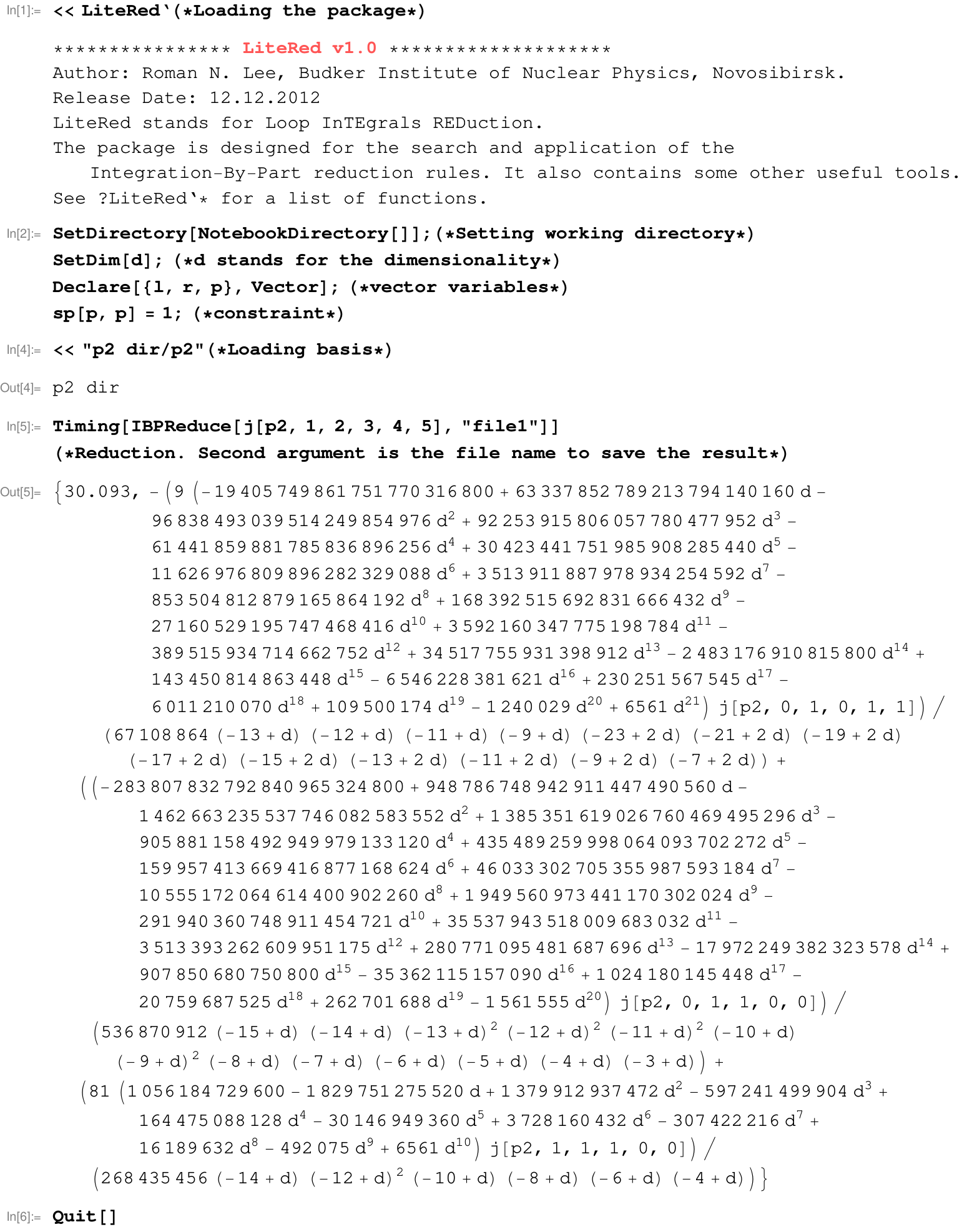}

\caption{A typical program of applying IBP rules.\label{fig:A-typical-program}}
\end{figure}

The central procedure which tries to construct the complete set of
symbolic rules for a given sector is \texttt{\textbf{SolvejSector}}
which we apply to each unique sector in this example. The result of
its work is the set of rules for each \texttt{\textbf{js{[}p2,...{]}}}
from the \texttt{\textbf{UniqueSectors{[}p2{]}}} list. All rules found
are saved in the directory ``p2 dir'' and ready to use for the reduction.
The typical program performing the reduction is shown in Fig. \ref{fig:A-typical-program}.

\subparagraph{Drawing graphs}

There is a possibility to draw graphs, corresponding to the integrals
and sectors. We remark that, in principle, the graph is determined,
up to some equivalences, by the set of internal lines. The possibility
to automatically determine the graph, corresponding to the set of
denominators, is planned in the future versions of the package. Meanwhile,
the present version implements the following. After defining the basis,
one can attach a graph to the highest sector(s) (which can be depicted
as a graph) by the command \texttt{\textbf{AttachGraph}}. Then the
graph for all subsectors is determined automatically.

For the above example there are no irreducible numerators, so the
sector \linebreak\texttt{\textbf{js{[}p2,1,1,1,1,1{]}}} can be depicted
by the graph. Then, the graph is attached by the command

\includegraphics{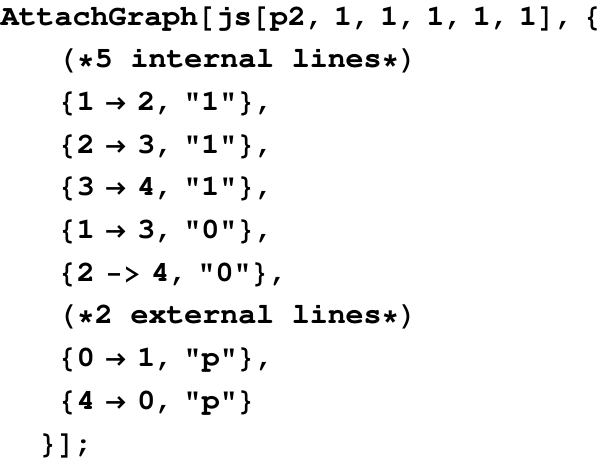}

Then, a graph of, say, sector\texttt{ }\texttt{\textbf{js{[}p2,1,1,1,1,0{]}}},
can be drawn with the command \texttt{\textbf{GraphPlot{[}jGraph{[}js{[}p2,1,1,1,1,0{]}{]}{]}}}.
Note that \texttt{\textbf{GraphPlot}} is a standard Mathematica function
and the presentation of the graph can be altered using its options,
as can be found in the examples distributed with the package.

\subparagraph{Additional tools}

Several additional tools are included in the package:
\begin{enumerate}
\item \texttt{\textbf{Dinv{[}j{[}...{]},sp{[}p,q{]}{]}}} \textbf{---} the
derivative with respect to the invariant constructed of the external
momenta, Eq. \eqref{eq:Dinv}.
\item \texttt{\textbf{RaisingDRR{[}basis,...{]}}}\textbf{ ---} the right-hand
side of the dimensional recurrence relation $j^{(d-2)}\left(\mathrm{basis},...\right)=\ldots$
, Eq. \eqref{eq:rDRR}. Note that the factor $\mu^{L}=-1$ for Minkovskian
metrics and odd number of loops should be taken into account manually.
\item \texttt{\textbf{LoweringDRR{[}basis,...{]}}}\textbf{ ---} the right-hand
side of the dimensional recurrence relation $j^{(d+2)}\left(\mathrm{basis},...\right)=\ldots$,
Eq. \eqref{eq:lDRR}. Note that the factor $\mu^{L}=-1$ for Minkovskian
metrics and odd number of loops should be taken into account manually.
\item \texttt{\textbf{FeynParUF{[}js{[}basis,...{]}{]}}} \textbf{--- }the
functions $U$ and $F$ entering the Feynman parametrization of the
integrals in the given sector.
\end{enumerate}

\subparagraph{Learning more}

There are several reduction examples in the directory \textbf{Examples}
of the archive file. One is encouraged to examine these examples for
some hints of the package usage. Another good starting point to know
more about the functions of the package is to submit a command \texttt{\textbf{?LiteRed`{*}}}.

\section{Implementation notes}

The \texttt{LiteRed} package depends on small packages \texttt{Types,
Numbers, Vectors, }and\texttt{ LinearFunctions}, which are also included
in the distributive. The\texttt{ Types} package allowas one to define
types and their transformation rules (e.g. vector plus vector is a
vector). The packages \texttt{Numbers, Vectors} introduce specific
types for number and vector variables. The package \texttt{LinearFunctions}
contains the function \texttt{\textbf{LFDistribute}} which distributes
linear functions over sum and pulls out the numbers and expressions
having type \texttt{\textbf{Number}}. 

\texttt{\textbf{AnalyzeSectors}} uses Criterion 1 from Ref. \cite{Lee2008}
for the determination of zero sectors. \texttt{\textbf{FindSymmetries}}\textbf{
}uses a combined approach based on the Feynman parametrization and
on the loop momenta shifts. In the first stage\texttt{\textbf{ FindSymmetries}}
finds mappings between the simple sectors using approach based on
Feynman parametrization very similar to the one described in Ref.
\cite{Pak2012}. This procedure works sufficiently fast even for complicated
examples. E.g., for the four-loop onshell mass operator topologies
the typical working time is a few minutes.

As it was stressed above, the package performs a heuristic search
of the reduction rules. The result of this search may strongly depend
on the order in which the functions $D_{\alpha}$ are listed, as well
as on the choice of the irreducible numerators. Therefore, in case
the program fails to find reduction rules, it makes sense to try changing
the irreducible numerators, as well as the listing order of $D_{\alpha}$.

\section{Conclusion}

In this short note we have presented a \textsl{Mathematica} package
\texttt{LiteRed} performing the IBP reduction of the multiloop integrals.
The package is based on the heuristic search of the reduction rules
(the procedure \texttt{\textbf{SolvejSector}}) and further application
of the rules found to the reduction problem. If the heuristic search
finishes successfully, the rules found present a very effective solution
of the reduction problem for the given class of the integrals. Similar
to the algorithm of costruction of s-bases in Ref. \cite{Smirnov:2006tz},
the heuristic search of the reduction rules is not proved to terminate
(and, in fact, seems to be not terminating in some complicated cases).
However, it appears that the search of the reduction rules, as implemented
in \texttt{LiteRed}, succeeds for a larger class of physically interesting
cases, see \textbf{Examples} folder in the distributive.

The package can be useful also for some other purposes. In particular,
the procedure \texttt{\textbf{FindSymmetries}} can be used to find
mapping between equivalent sectors.

\acknowledgments I am very grateful to Vladimir and Alexander Smirnovs
for valuable remarks and discussions. I highly appreciate warm hospitality
and financial support of TTP KIT, Karlsruhe, and ITP UZH, Zurich,
where a part of this work was done. This work is supported by the
Russian Foundation for Basic Research through grant 11-02-01196 and
by the Ministry of Education and Science of the Russian Federation. 

\appendix
\bibliographystyle{JHEP}
\bibliography{LiteRed}

\end{document}